\documentclass[12pt]{article}
\usepackage{amsmath}
\usepackage{amssymb}
\usepackage{array}
\usepackage{geometry}
\usepackage{graphicx}
\usepackage{enumerate}
\usepackage[small, compact] {titlesec}
 \numberwithin{equation}{section}
\begin{document}
\vskip7cm\noindent
\begin{center}{\bf STOCHASTIC PROCESSES VIA THE PATHWAY MODEL}\\ 
\vskip.3cm{\bf Arak M. Mathai}\\
\vskip.2cm{Centre for Mathematical and Statistical Sciences, Peechi Campus, KFRI}\\
\vskip.1cm{Peechi-680653, Kerala, India}\\
directorcms458@gmail.com\\
\vskip.2cm{and}\\
\vskip.2cm{Department of Mathematics and Statistics, McGill University}\\
\vskip.1cm{805 Sherbrooke Street West, Montreal, Quebec, Canada, H3A2K6}\\

\vskip.2cm{\bf Hans J. Haubold}\\
\vskip.2cm{Centre for Mathematical and Statistical Sciences, Peechi Campus, KFRI}\\
\vskip.1cm{Peechi-680653, Kerala, India}\\
hans.haubold@gmail.com\\
\vskip.2cm{and}\\
\vskip.1cm{Office for Outer Space Affairs, United Nations}\\
\vskip.1cm{P.O. Box 500, Vienna International Center, A-1400 Vienna, Austria}\\
 
\end{center}

\vskip.5cm\noindent{\bf Abstract} \vskip.3cm After collecting data from
observations or experiments, the next step is to build an appropriate mathematical or stochastic model to describe
the data so that further studies can be done with the help of the models. In this article, the input-output type mechanism is considered
first, where reaction, diffusion, reaction-diffusion, and production-destruction type physical situations can fit
in. Then techniques are described to produce thicker or thinner tails (power law behavior) in stochastic models. Then the pathway idea is
described where one can switch to different functional forms of the probability density function) through a parameter called the pathway parameter.



\vskip.5cm\noindent{\bf 1.\hskip.3cm Introduction}
\vskip.3cm After collecting data from experiments or from observations, the next step is to study the data and make
inference out of the data. This can be achieved by using mathematical methods and models if the physical situation is deterministic
in nature, otherwise create stochastic models if the physical situation is non-deterministic in nature. If the
underlying phenomenon, which created the data, is unknown, possibly deterministic but the underlying factors and the
way in which these factors act are unknown, thereby the situation becomes random in nature. Then we go for stochastic
models or non-deterministic type models. One has to know or speculate about the underlying factors as well as the way
in which these factors act so that one can decide which category of models are appropriate. If the observations are
available over time then a time series type of model may be appropriate. If the time series shows periodicities then
each cycle can be analyzed by using specific types of stochastic models.
\vskip.2cm Here we will consider models to
describe short-term behavior of data or behavior within one cycle if a cyclic behavior is noted. When monitoring solar
neutrinos it is seen that there is likely to be an eleven year cycle
 and within each cycle the behavior of the graph is something like slow increase with several local peaks to a
 maximum peak and slow decrease with humps back to normal level. In such situations,  what is observed is not really 
what is actually produced. What is observed
 is the residual part of what is produced minus what is consumed or converted and thus the actual observation
 is made on the residual part only. Many of natural phenomena belong to this type of behavior of the form
 $u=x-y$ where $x$ is the input or production variable and $y$ is the output or consumption or destruction
 variable and $u$ represents the residual part which is observed. A general analysis of input-output situation may be seen from [21]. In many situations one can assume
  that $x$ and $y$ are statistically independently distributed and that $u\ge 0$ means production dominates over
  destruction or input dominates over the output.
  \vskip.2cm In reaction-rate theory, when particles react with each other producing new particles or producing
  neutrinos we may have the following type of situations. Certain particles may react with each other in short-span
  or short-time periods and produce small number of particles, others may take medium time intervals and produce
  larger numbers of particles and yet others may react over a long span and produce larger number of particles.
  For describing such types of situations in the production of solar neutrinos the present authors considered
  creating mathematical models by erecting triangles whose ares are proportional to the neutrinos produced, see [5],[6],[7],[8].

  \vskip.2cm Another approach that was adopted was to assume $x$ and $y$ as independently distributed random
  variables, then work out the density of the residual variable under the assumption that $x-y\ge 0$. The simplest such situation is an exponential type input and an exponential type output. Then we can look at sum of such independently distributed residual type variables. This is a reasonable type of assumption.  Then input-output model has the Laplace density, when $x$ and $y$ are identically and independently distributed and the density is given by,

  $$f_1(u)=\frac{\beta_1}{2}{\rm e}^{-\beta_1|u-\alpha_1|},0\le
  u<\infty,\beta_1>0,\eqno(1.1)
  $$and $f_1(u)=0$ elsewhere, where $\alpha_1$ is a location parameter. Note that $\beta_1$ can act as a scale parameter or as a dispersion or scatter parameter. Suppose that this situation is repeated at successive locations and with the scale parameter $\beta=\beta_1,\beta_2,...$.  Then the nature of the graph will be that of a sum of Laplace densities. If the location parameters are sufficiently farther apart then the graph will look like that in Figure 1(b). If such blips are occurring sufficiently close together then we have a graph of the type in Figure 1(a). In these graphs we have taken only five to six locations for simplicity. But by taking successive locations we can generate many of the phenomena that are seen in nature, especially in time series data. When the locations are sufficiently closer we get the graph with several local maxima/spikes and a continuous curve. This is the type of behavior seen in solar neutrino production. Cyclic patterns can also arise depending upon the location and scale parameters. Here $\beta_1 $ measures the intensity of the blip and $\alpha_1$ the location where it happens, and each blip is the residual effect of an exponential type input and an independent exponential type output of the same strength. If $\alpha_1,\alpha_2,...$ are farther apart then the contributions coming from other blips will be negligible and if $\alpha_1,\alpha_2,...$ are close together then there will be contributions from other blips. The function will be of the following form:
  $$f(u)=\sum_{j=1}^k\frac{\beta_j}{2}{\rm e}^{-\beta_j|u-\alpha_j|}, 0\le u<\infty, \beta_j>0, j=1,...,k<\infty\eqno(1.2)
  $$If one requires $f(u)$ to be a density within a number of spikes then divide the sum by $k$ so that we have a convex combination of Laplace densities, which will again be a density. The model does not require that we create a density out of the pattern. If the arrival of the location points ($\alpha_j$) is governed by a Poisson process then we will have a Poisson mixture of Laplace densities.

  \begin{center}
{\includegraphics[width=6cm, height=5cm]{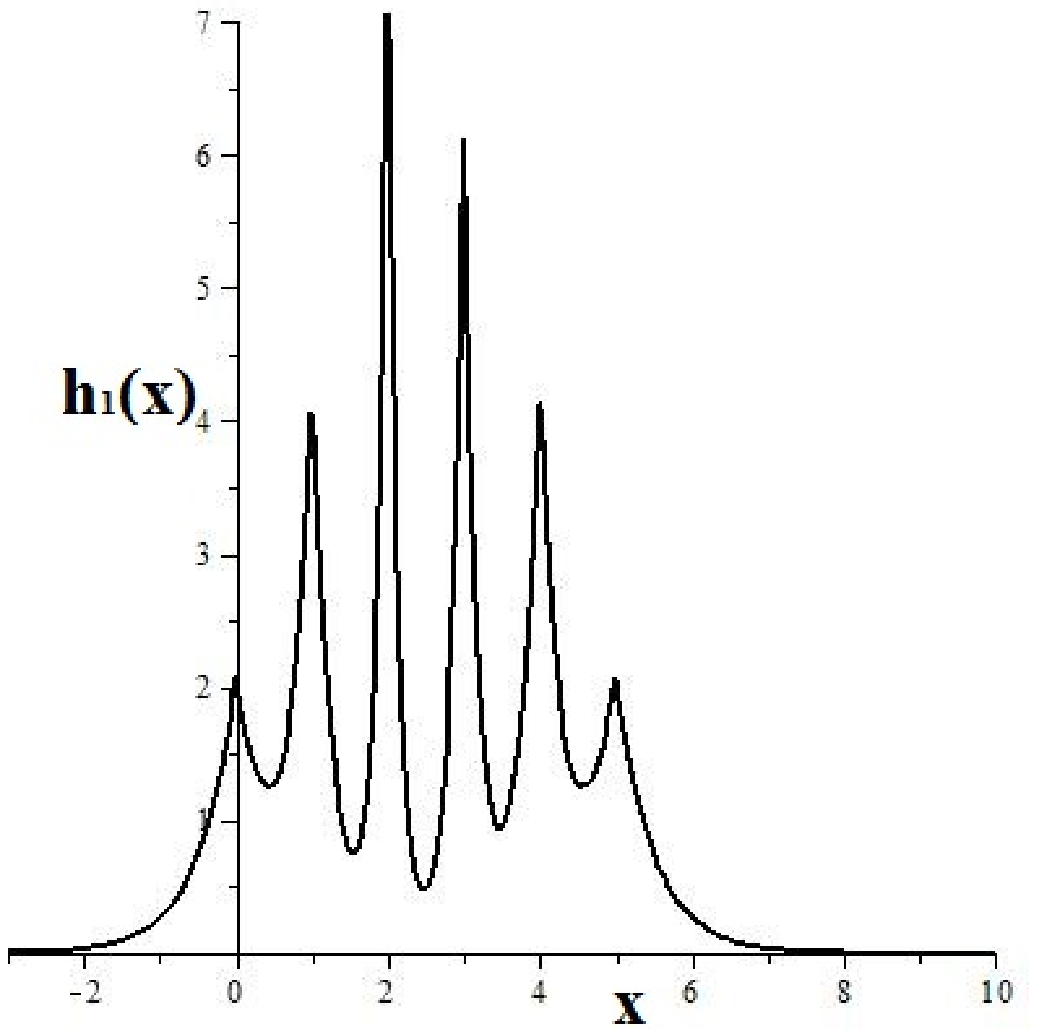}}(a)
{\includegraphics[width=6cm, height=5cm]{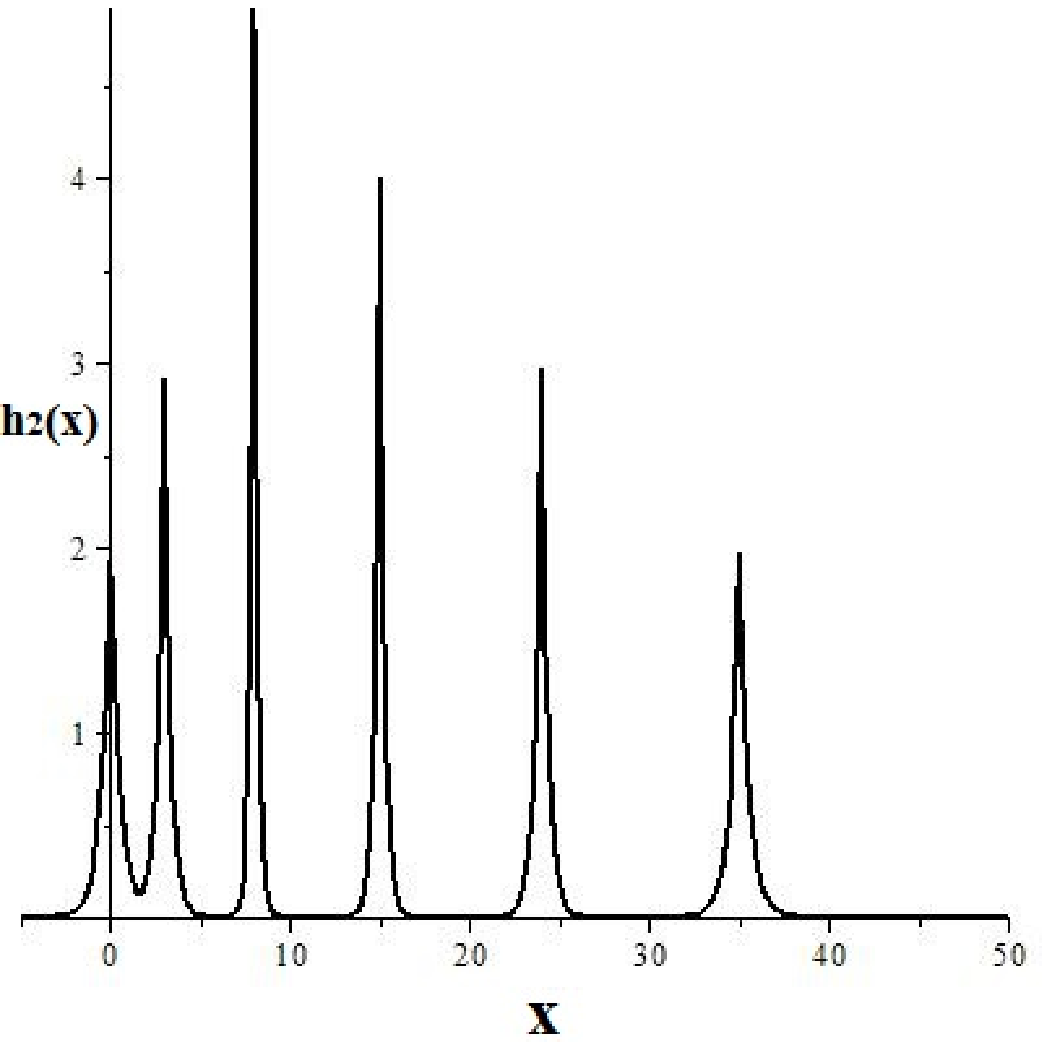}}(b)
\end{center}

 \noindent \begin{center}Figure 1 (a)\hskip1.5cm (b)\end{center}
  \vskip.3cm A symmetric Laplace density will be of the following form:

  $$f_2(u)=\frac{1}{2\beta}{\rm e}^{-\frac{|u|}{\beta}},~-\infty<u<\infty\eqno(1.3)
  $$and the graph is of the following form:
  \begin{center}
{\includegraphics[width=12cm, height=6cm]{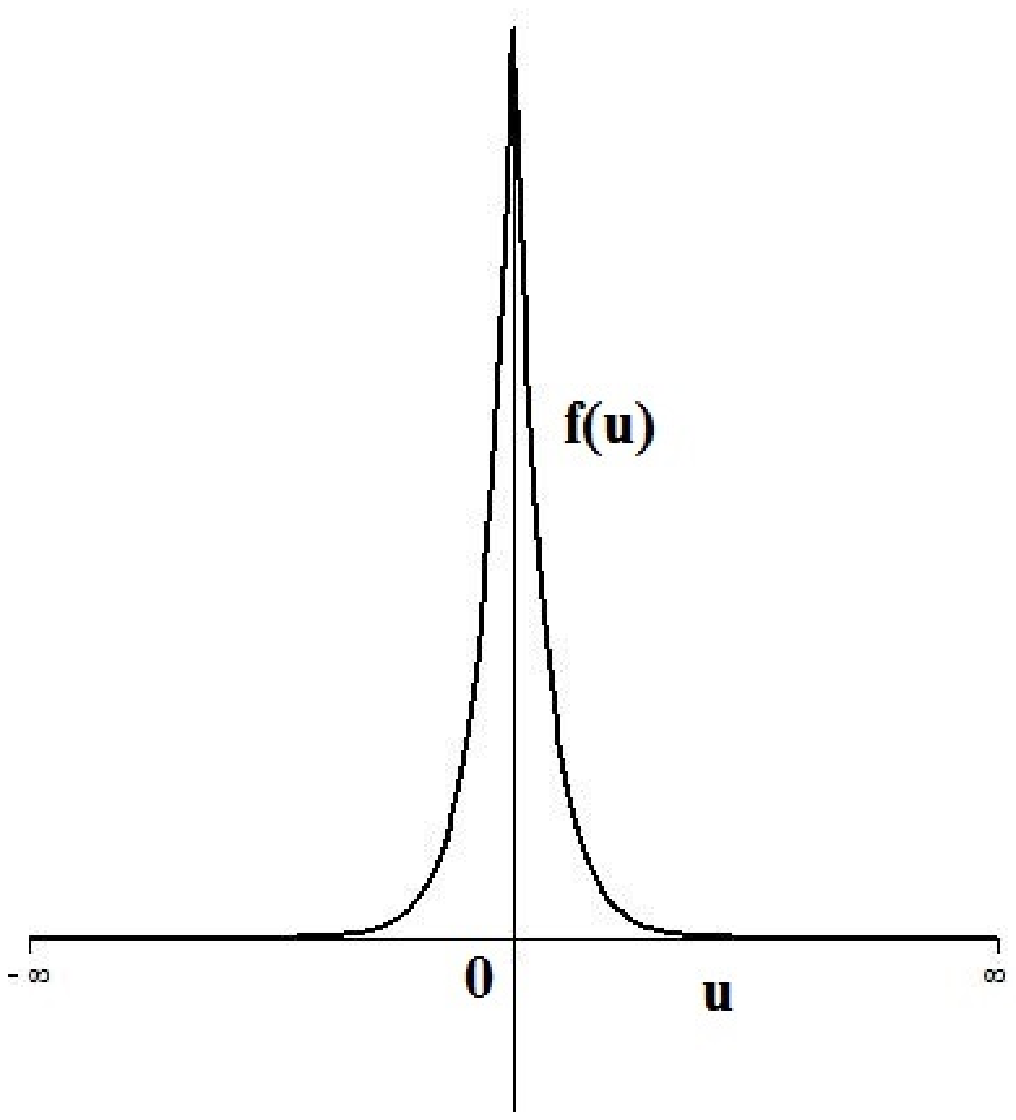}}\\
Figure 2~~Symmetric Laplace density
\end{center}
  \vskip.3cm This is the symmetric case where $u<0$ behaves the same way as $u\ge 0$. If the behavior of $u$ is different for $u<0$ and $u\ge 0$ then we get the asymmetric Laplace case which can be written as
 $$ g(u)=\begin{cases}\frac{1}{(\beta_1+\beta_2)}{\rm e}^{\frac{u}{\beta_1}},~-\infty<u<0\\
  \frac{1}{(\beta_1+\beta_2)}{\rm e}^{-\frac{u}{\beta_2}},~0\le u<\infty
  \end{cases}\eqno(1.4)$$
  \begin{center}
{\includegraphics[width=12cm, height=6cm]{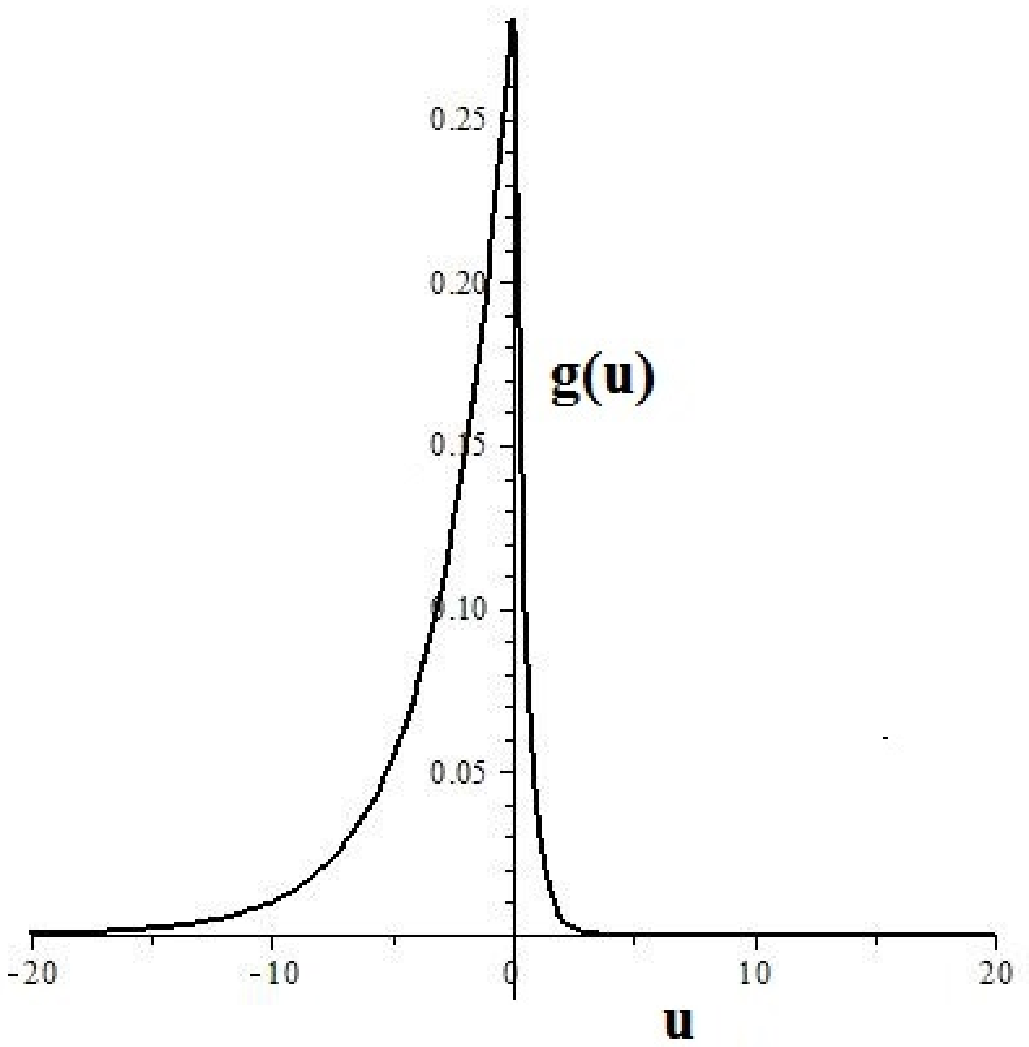}}\\
Figure 3~~Asymmetric Laplace case
\end{center}
  \vskip.3cm When $\alpha_1>1,\alpha_2>1,\alpha_1=\alpha_2=\alpha,\beta_1=\beta_2=\beta$ we have independently and identically distributed gamma random variables for $x$ and $y$ and $u=x-y$ is the difference between them. Then $g_1(u)$ can be seen to be the following:

  $$g_1(u)=\frac{u^{2\alpha-1}{\rm e}^{-\frac{u}{\beta}}}{\beta^{2\alpha}\Gamma^2(\alpha)}\int_{z=0}^{\infty}(1+z)^{\alpha-1}z^{\alpha-1}{\rm e}^{-\frac{1}{\beta}(2uz)}{\rm d}z\eqno(1.5)
  $$for $u\ge 0$, $\alpha>0,\beta>0$. This behaves like a gamma density and provides a symmetric model for $u\ge 0$ and $u<0$. The following is the nature of the graph.
\begin{center}
{\includegraphics[width=12cm, height=6cm]{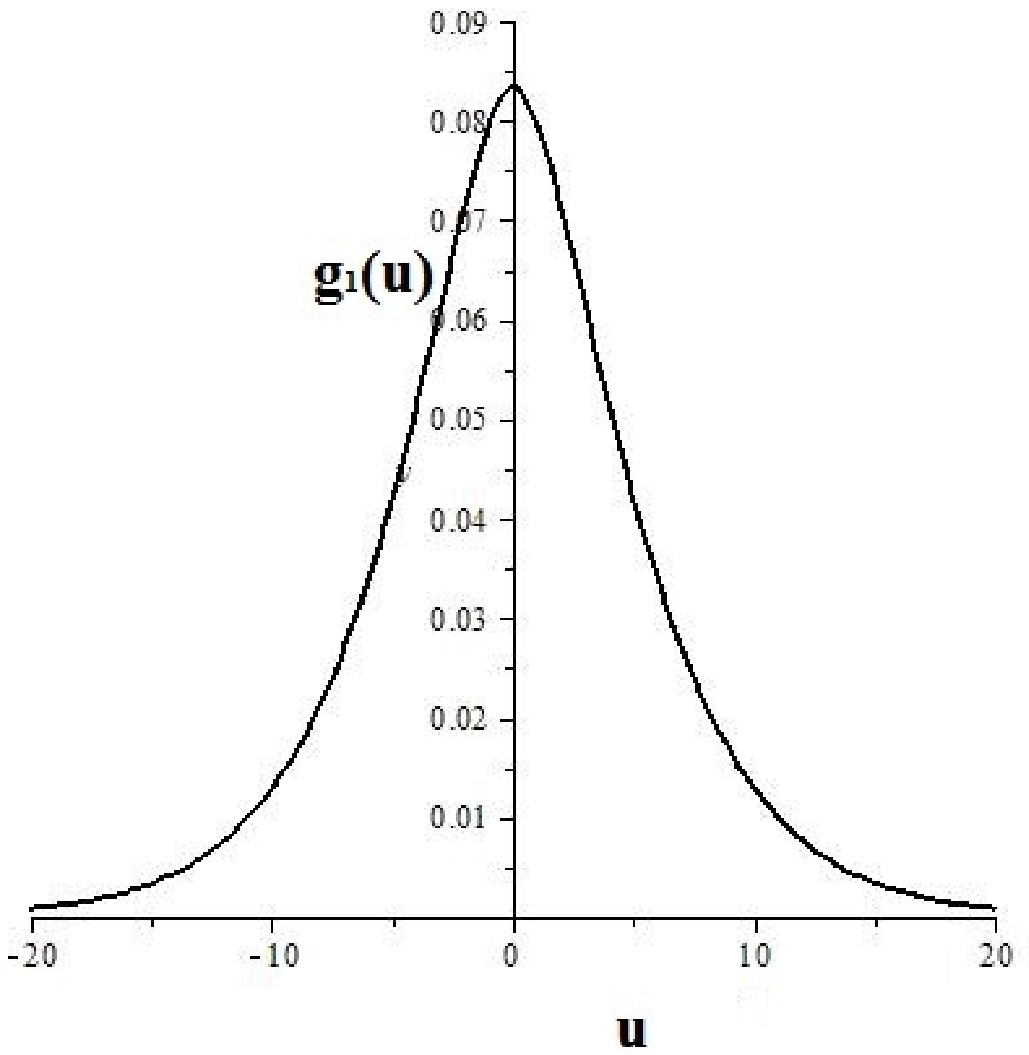}}\\
Figure 4~~~$g_1(u)$ in the symmetric gamma type input-output variables
\end{center}

  \vskip.3cm\noindent{\bf 2.\hskip.3cm Models with Thicker and Thinner Tails}
  \vskip.3cm For a large number of situations a gamma type model may be appropriate. A two parameter gamma density is of the type

  $$f(x)=\frac{1}{\beta^{\alpha}\Gamma(\alpha)}x^{\alpha-1}{\rm e}^{-\frac{x}{\beta}},x\ge 0,\alpha>0,\beta>0.\eqno(2.1)
  $$Sometimes a member from this parametric family of functions may be appropriate to describe a data set. Sometimes the data require a slightly thicker-tailed model due to chances of higher probabilities or more area under the curve in the tail. Two of such models developed by the authors' groups will be described here. One type is where the model in (2.1) is appended with a Mittag-Leffler series and another type is where (2.1) is appended with a Bessel series, see also [25].

  \vskip.3cm\noindent{\bf 2.1.\hskip.3cm Gamma model with appended Mittag-Leffler function}
  \vskip.3cm Consider a gamma density of the type

  $$g_3(x)=c_1~x^{\gamma-1}{\rm e}^{-\frac{x}{\delta}},~\delta>0,~\gamma>0,~x\ge 0.
  $$Suppose that we append this $g_3(x)$ with Mittag-Leffler function $E_{\alpha,\gamma}^{\beta}(-x^{\alpha})$ where
  $$E_{\alpha,\gamma}^{\beta}(-ax^{\alpha})=\sum_{k=0}^{\infty}\frac{(\beta)_k}{k!}(-a)^k\frac{{x^{\alpha}}^k} {\Gamma(\gamma+\alpha k)},~\alpha>0,~\gamma>0.
  $$Consider the function

  $$f^{*}(x)=c~\sum_{k=0}^{\infty}\frac{(\beta)_k}{k!}(-a)^k\frac{x^{\alpha k+\gamma-1}{\rm e}^{-\frac{x}{\delta}}}{\Gamma(\gamma+\alpha k)},~x\ge 0
  $$where $c$ is the normalizing constant. Let us evaluate $c$. Since the total integral is $1$,
  \begin{align}
  1&=\int_0^{\infty}f^{*}(x){\rm d}x=c\sum_{k=0}^{\infty}\frac{(\beta)_k}{k!}(-a)^k\int_0^{\infty}\frac{x^{\alpha k+\gamma-1}{\rm e}^{-\frac{x}{\delta}}}{\Gamma(\gamma+\alpha k)}{\rm d}x\nonumber\\
  &=c\sum_{k=0}^{\infty}\frac{(\beta)_k}{k!}(-a)^k\delta^{\alpha k+\gamma}=c~\delta^{\gamma}(1+a\delta^{\alpha})^{-\beta},~|a\delta^{\alpha}|<1\nonumber
  \end{align}for $\beta>0,\alpha>0,\delta>0,a\beta \delta^{\alpha}<1, |a\delta^{\alpha}|<1.$ Therefore the density is
  $$f^{*}(x)=\frac{(1+a\delta^{\alpha})^{\beta}}{\delta^{\gamma}}x^{\gamma-1}{\rm e}^{-\frac{x}{\delta}}\sum_{k=0}^{\infty}\frac{(\beta)_k}{k!}\frac{(-a)^k\delta^{\alpha k}}{\Gamma(\gamma+\alpha k)}
  $$for $0\le x<\infty,\alpha>0,\gamma>0,\delta>0,\beta>0$, $|a\delta^{\alpha}|<1,a\beta \delta^{\alpha}<1$. That is,
  $$f^{*}(x)=\frac{(1+a\delta^{\alpha})^{\beta}}{\delta^{\gamma}}x^{\gamma-1}{\rm e}^{-\frac{x}{\delta}}[\frac{1}{\Gamma(\gamma)}+\sum_{k=1}^{\infty}\frac{(\beta)_k}{k!}\frac{(-1)^k\delta^{\alpha k}}{\Gamma(\gamma+\alpha k)}].
  $$Note that $a=0$ corresponds to the original gamma density. The following are some graphs of the appended Mittag-Leffler-gamma density. When $a<0$ we have thinner tail and when $a>0$ we have thicker tails compared to the gamma tail.
\begin{center}
{\includegraphics[width=12cm, height=6cm]{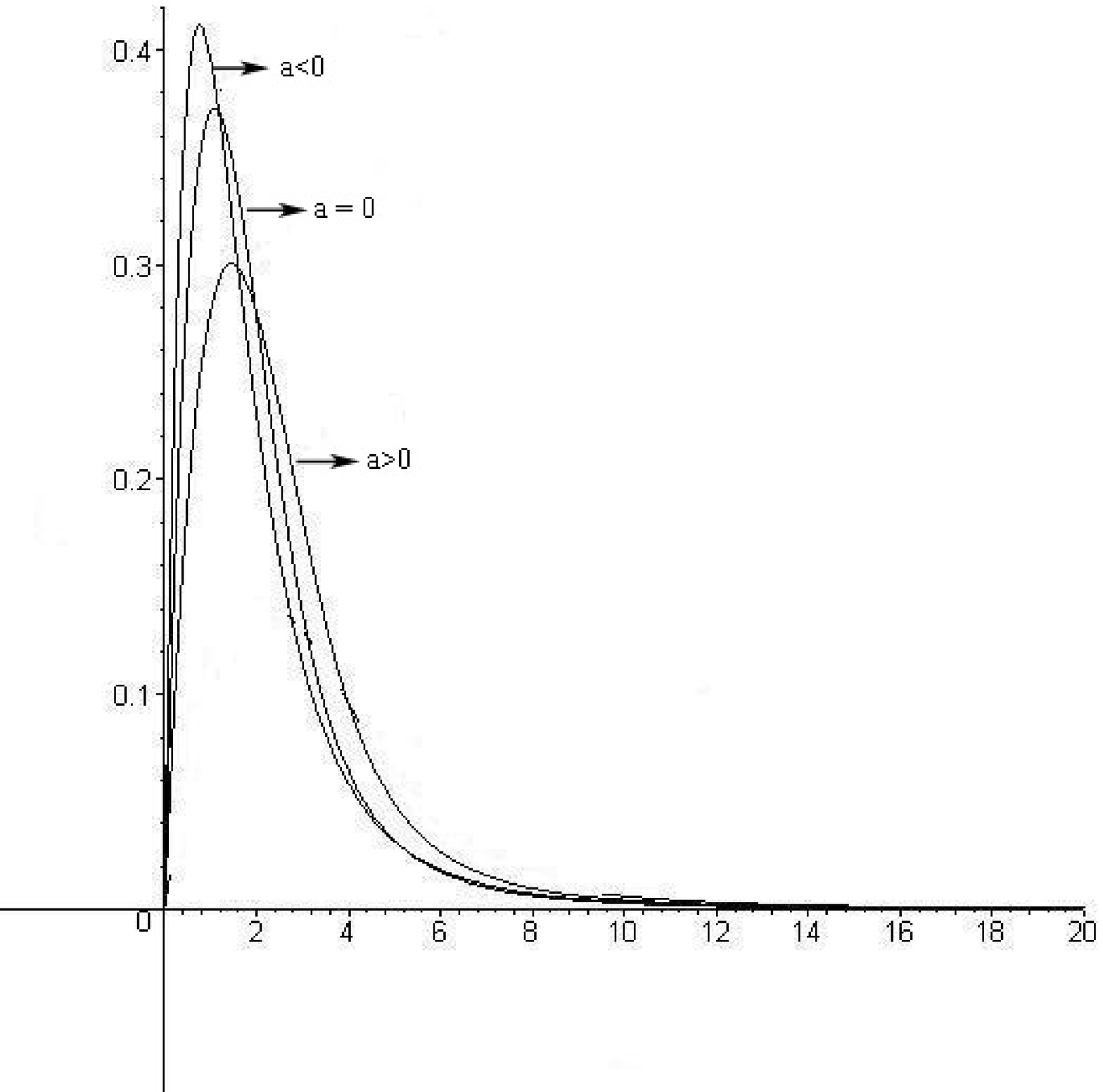}}\\
Figure 5~Gamma density with Mittag-Leffler function appended
\end{center}

  \vskip.3cm\noindent{\bf 2.2.\hskip.3cm Bessel appended gamma density}

  \vskip.3cm Consider the model of the type of a basic gamma density appended with a Bessel function, see also [25].

  $$\tilde{f}(x)=c~x^{\gamma-1}{\rm e}^{-\frac{x}{\delta}}\sum_{k=0}^{\infty}\frac{x^k(-a)^k}{k!\Gamma(\gamma+k)},~\delta>0,\gamma>0,x\ge 0,
  $$where $c$ is the normalizing constant. The appended function is of the form
  $$\frac{1}{\Gamma(\gamma)}{_0F_1}(~~;\gamma :-ax)
  $$which is a Bessel function. Let us evaluate $c$.
  \begin{align}
  1&=c~\sum_{k=0}^{\infty}\frac{(-a)^k}{k!}\int_0^{\infty}\frac{x^{\gamma+k-1}}{\Gamma(\gamma+k)}{\rm e}^{-\frac{x}{\delta}}{\rm d}x\nonumber\\
  &=c~\delta^{\gamma}\sum_{k=0}^{\infty}\frac{(-a)^k\delta^k}{k!}=c~\delta^{\gamma}{\rm e}^{-a\delta}.\nonumber
  \end{align}Hence the density is of the form

  $$\tilde{f}(x)=\frac{{\rm e}^{a\delta}}{\delta^{\gamma}}x^{\gamma-1}{\rm e}^{-\frac{x}{\delta}}\sum_{k=0}^{\infty}\frac{x^k(-a)^k}{k!\Gamma(\gamma+k)},x\ge 0,~\gamma>0,\delta>0.
  $$
  \begin{center}
{\includegraphics[width=12cm, height=6cm]{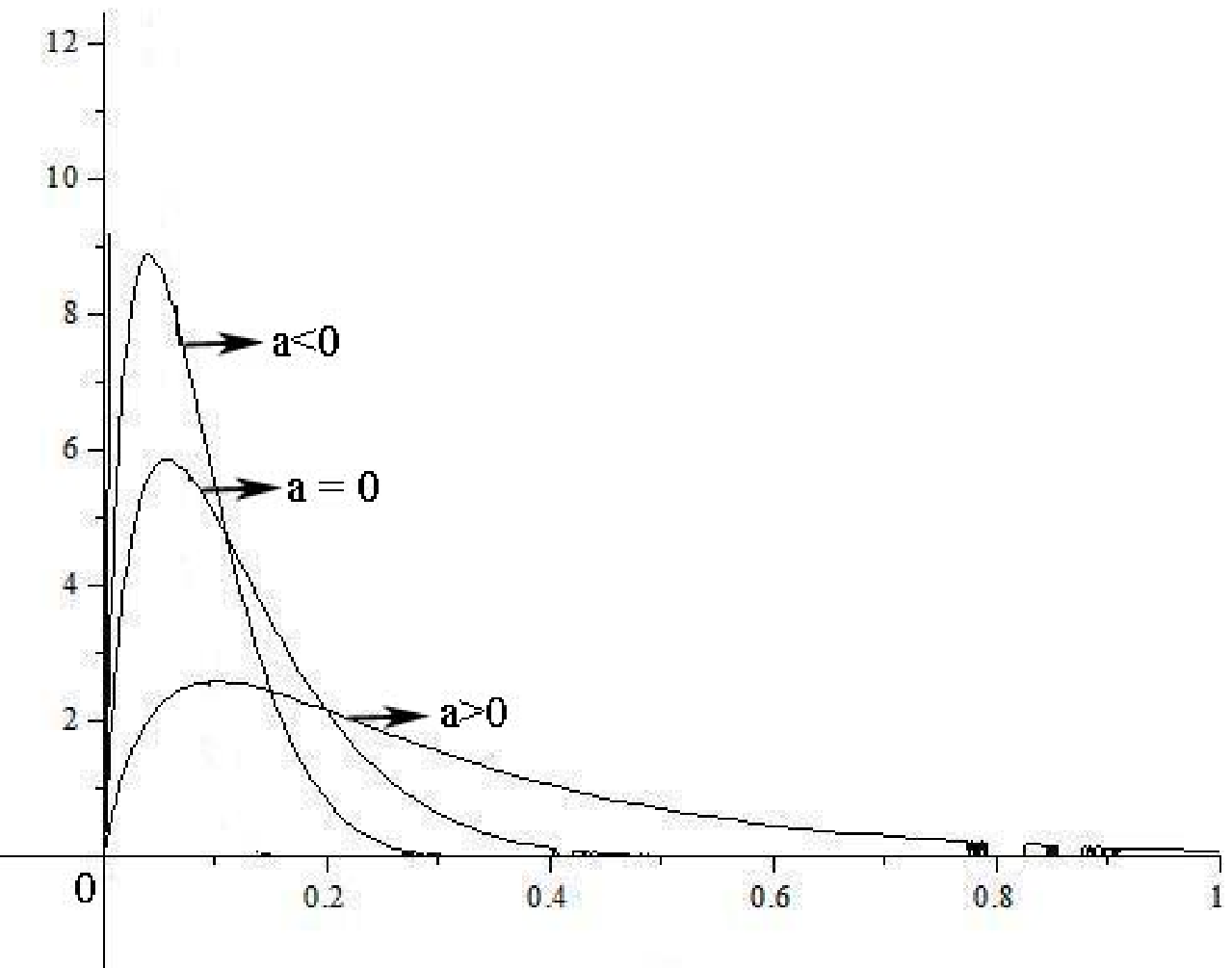}}\\
Figure 6~~ Gamma appended with Bessel function
\end{center}

  \vskip.3cm\noindent{\bf Note}:\hskip.3cm Instead of appending with Bessel function one could have appended with a general hypergeometric series. But a general hypergeometric series does not simplify into a convenient form. We have chosen specialized parameters as well as suitable functions so that the normalizing constants simplify to convenient forms thereby general computations will be much easier and simpler.

  \vskip.3cm\noindent {\bf 3.\hskip.3cm Pathway Idea}
  \vskip.3cm Here we consider a model which can switch to three functional forms covering almost all statistical densities in current use, see [23]. Let

  $$f_1^{*}(x)=c_1^{*}|x|^{\gamma}[1-a(1-\alpha)|x|^{\delta})^{\frac{\eta}{1-\alpha}},\alpha<1,\eta>0,a>0,\delta>0,\eqno(3.1)
  $$and $1-a(1-\alpha|x|^{\delta}>0$, where $c_1^{*}$ is the normalizing constant. When $\alpha<1$ the model in (3.1) stays as the generalized type-1 beta family, extended over the real line. When $\alpha>1$ write $1-\alpha=-(\alpha-1)$ with $\alpha>1$. Then the functional form in (3.1) changes to

  $$f_2^{*}(x)=c_2^{*}|x|^{\gamma}[1+a(\alpha-1)|x|^{\delta}]^{-\frac{\eta}{\alpha-1}}\eqno(3.2)
  $$for $\alpha>1,a>0,\eta>0,-\infty<x<\infty$. Note that (3.2) is the extended generalized type-2 beta family of functions. When $\alpha\to 1$ then both (3.1) and (3.2) go to

  $$f_3^{*}(x)=c_3^{*}~|x|^{\gamma}{\rm e}^{-a\eta|x|^{\delta}},a>0,\eta>0,\delta>0,-\infty<x<\infty.\eqno(3.3)
  $$Eq. (3.3) is the extended generalized gamma family of functions. Thus (3.1) is capable of switching to three families of functions. This is the pathway idea and $\alpha$ is the pathway parameter. Through this parameter $\alpha$ one can reach the three families of functions in (3.1),(3.2), and (3.3). The pathway idea was introduced by Mathai [23]. The normalizing constants can be seen to be the following:
  $$c_1^{*}=\frac{\delta}{2}\frac{[a(1-\alpha)]^{\frac{\gamma+1}{\delta}}\Gamma(\frac{\gamma+1}{\delta}+\frac{\eta}{1-\alpha}+1)}
  {\Gamma(\frac{\gamma+1}{\delta})\Gamma(\frac{\eta}{1-\alpha}+1)}\eqno(3.4)
  $$for $a>0,\alpha<1,\delta>0,\gamma>-1,\eta>0$,
  $$c_2^{*}=\frac{\delta}{2}\frac{[a(\alpha-1)]^{\frac{\gamma+1}{\delta}}\Gamma(\frac{\eta}{\alpha-1})}
  {\Gamma(\frac{\gamma+1}{\delta})\Gamma(\frac{\eta}{\alpha-1}-\frac{\gamma+1}{\delta})}\eqno(3.5)
  $$for $\alpha>1,a>0,\delta>0,\eta>0,\delta>0,\frac{\eta}{\alpha-1}-\frac{\gamma+1}{\delta}>0$,
  $$c_3^{*}=\frac{\delta}{2}\frac{(a\eta)^{\frac{\gamma+1}{\delta}}}{\Gamma(\frac{\gamma+1}{\delta})},a>0,\delta>0,\eta>0,
  \gamma>-1.\eqno(3.6)
  $$Note that (3.1) is a finite range model, suitable to describe situations where the tails are cut off. When $\alpha$ comes closer and closer to $1$ then the cut-off point moves away from the origin and eventually goes to $\pm \infty$. When $\alpha\to 1$ then model (3.1) goes to model (3.3) which is an extended generalized gamma model. The model in (3.2) is type-2 beta form, spreads out over the whole real line and the shape will be closer to that of a gamma type model when $\alpha$ approaches $1$. Thus the pathway models in (3.1),(3.2), and (3.3) cover all types of situations where the tails are cut off, tails are made thinner or thicker compared to a gamma type model. The extended gamma type model in (3.3) also contains the Gaussian model, Brownian motion, Maxwell-Boltzmann density etc. If Gaussian or Maxwell-Boltzmann is the stable or ideal form in a physical situation then the unstable neighborhoods are covered by (3.1) and (3.2) or the paths leading to this stable form is described by (3.1) and (3.2).
  \vskip.2cm It is worth noting that (3.1) for $x>0,\gamma=0,a=1,\delta=1,\eta=1$ is the Tsallis statistics for non-extensive statistical mechanics. Also note that (3.2) for $a=1,\delta=1,\eta=1$ is superstatistics. This superstatistics can also be derived as the unconditional density when both the conditional density of $x$ given a parameter $\theta$ and the marginal density of $\theta$ are gamma densities or exponential type densities, the details may be seen from Mathai and Haubold ([15], [26], [27], [29], [30]).
  \vskip.2cm Various types of models which are applicable in a variety of situations may be seen from Mathai [25].

  \vskip.3cm\noindent{\bf 4.\hskip.3cm Reaction Rate Probability Integral Model}

  \vskip.3cm Starting from 1980's the present authors had pursued mathematical models for reaction-rate theory in various situations such as non-resonant reactions and resonant reactions under various cases such as depletion, high energy tail cut off etc, see [2],[3],[4], [5], [9], [11]. The basic model is an integral of the following form:
  $$I_{(1)}=\int_0^{\infty}x^{\gamma-1}{\rm e}^{-ax^{\delta}-zx^{-\rho}},a>0,z>0,\rho>0,\delta>0.\eqno(4.1)
  $$For $\rho=\frac{1}{2}, \delta=1$ one has the basic probability integral in the non-resonant case, see [4]. For $\gamma=0,\rho=1$ one has Kr\"atzel integral [24]. For $\gamma=0,\delta=1,\rho=1$ one has inverse Gaussian density. Computational aspect of (4.1) is discussed in [1] and related material may be seen from [20]. Since the integral in (4.1) is a product of integrable functions one can evaluate the integral in (4.1) with the help of Mellin convolution of a product because the integrand can be written as
  $$\int_0^{\infty}\frac{1}{v}f_1(v)f_2(\frac{u}{v}){\rm d}v,f_1(x)=x^{\gamma}{\rm e}^{-ax^{\delta}},f_2(y)={\rm e}^{-y^{\rho}}\eqno(4.2)
  $$for $u=z^{\frac{1}{\rho}},~u=xy$. Then the Mellin convolution of the integral in (4.1), denoting the Mellin transform of a function $f$ with Mellin parameter $s$ as $M_f(s)$, we have from (4.1)
  $$M_{I_{(1)}}(s)=M_{f_1}(s)M_{f_2}(s),\eqno(4.3)
  $$where
  \begin{align}
  M_{f_1}(s)&=\int_0^{\infty}x^{s-1}f_1(x){\rm d}x=\int_0^{\infty}x^{\gamma+s-1}{\rm e}^{-ax^{\delta}}{\rm d}x\nonumber\\
  &=\frac{1}{\delta}\frac{\Gamma(\frac{s+\gamma}{\delta})}{a^\frac{s+\gamma}{\delta}},\Re(s+\gamma)>0\nonumber
  \end{align}and

  $$M_{f_2}(s)=\int_0^{\infty}y^{s-1}{\rm e}^{-y^{\rho}}{\rm d}y=\frac{1}{\rho}\Gamma(\frac{s}{\rho}),~\Re(s)>0.
  $$Hence
  $$M_{I_{(1)}}(s)=M_{f_1}(s)M_{f_2}(s)=\frac{1}{\rho\delta~a^{\frac{\gamma}{\delta}}}\frac{\Gamma(\frac{s+\gamma}{\delta})\Gamma(\frac{s}{\rho})}{a^{\frac{s}{\delta}}}.
  $$Therefore the integral in (4.1) is the inverse Mellin transform of (4.3). That is,
  \begin{align}
  I_{(1)}&=\frac{1}{2\pi i}\int_{c-i\infty}^{c+i\infty}\frac{1}{\rho\delta a^{\frac{\gamma}{\delta}}}\Gamma(\frac{s+\gamma}{\delta})\Gamma(\frac{s}{\rho})(ua^{\frac{1}{\delta}})^{-s}{\rm d}s,u=z^{\frac{1}{\rho}},i=\sqrt{-1}\nonumber\\
  &=\frac{1}{\rho\delta a^{\frac{\gamma}{\delta}}}H_{0,2}^{2,0}[z^{\frac{1}{\rho}}a^{\frac{1}{\delta}}\big\vert_{(0,\frac{1}{\rho}),
  (\frac{\gamma}{\delta},\frac{1}{\delta})}]&(4.4)\nonumber
  \end{align}where $H(\cdot)$ is the H-function, see [28],[32]. From the basic result in (4.4) we can evaluate the reaction-rate probability integrals in the other cases of non-relativistic reactions.

  \vskip.3cm\noindent{\bf 4.1.\hskip.3cm Generalization of reaction-rate models}
  \vskip.3cm A companion integral corresponding to (4.1) is the integral
  $$I_{(2)}=\int_0^{\infty}x^{\gamma}{\rm e}^{-ax^{\delta}-zx^{\rho}}{\rm d}x, a>0,\delta>0,\rho>0,z>0.\eqno(4.5)
  $$In (4.1) we had $x^{-\rho}$ with $\rho>0$ whereas in (4.5) we have $x^{\rho}$ with $\rho>0$. For $\delta=1$, (4.5) corresponds to the Laplace transform or moment generating function of a generalized gamma density in statistical distribution theory. The integral in (4.5) can be evaluated. Note that (4.5) can be written in the form of an integral of the form
  $$\int_0^{\infty}vf_1(v)f_2(uv){\rm d}v,f_1(x)=x^{\gamma-1}{\rm e}^{-ax^{\delta}},f_2(y)={\rm e}^{-y^{\rho}}\eqno(4.6)
  $$for $u=z^{\frac{1}{\rho}}$. The integral in (4.6) is in the structure of a Mellin transform of a ratio $u=\frac{y}{x}$, so that the Mellin transform of $I_{(2)}$ is then
  $$M_{I_{(2)}}(s)=M_{f_2}(s)M_{f_1}(2-s).\eqno(4.7)
  $$The inverse Mellin transform in (4.7) gives the integral $I_{(2)}$. The pair of integrals $I_{(1)}$ and $I_{(2)}$ belong to a particular case of a general versatile model considered by the authors earlier [33].
  \vskip.2cm A generalization of $I_{(1)}$ and $I_{(2)}$ is the pathway generalized model, which results in the versatile integral. The pathway generalization is done by replacing the two exponential functions by the corresponding pathway form. Consider the integrals of the following types:

  $$I_p=\int_0^{\infty}x^{\gamma}[1+a(q_1-1)x^{\delta}]^{-\frac{1}{q_1-1}}[1+b(q_2-1)x^{\rho}]^{-\frac{1}{q_2-1}}{\rm d}x,\eqno(4.8)
  $$where $q_1>1,q_2>1,a>0,b>0,$. We will keep $\rho$ free, could be negative or positive. Note that
  $$\lim_{q_1\to 1}[1+a(q_1-1)x^{\delta}]^{-\frac{1}{q_1-1}}={\rm e}^{-ax^{\delta}}
  $$and
  $$\lim_{q_2\to 1}[1+b(q_2-1)x^{\rho}]^{-\frac{1}{q_2-1}}={\rm e}^{-bx^{\rho}}.
  $$Hence
  $$\lim_{q_1\to 1,q_2\to 1}I_p=\int_0^{\infty}x^{\gamma}{\rm e}^{-ax^{\delta}-bx^{\rho}}{\rm d}x
  $$which is the integral in (4.5) and if $\rho<0$ then it is the integral in (4.1). The general integral in (4.8) belongs to the general family of versatile integrals. The factors in the integrand in (4.8) are of the generalized type-2 beta form. We could have taken each factor in type-1 beta or type-2 beta form, thus providing 6 different combinations. For each case, we could have the situation of $\rho>0$ or $\rho<0$. The whole collection of such models is known as the versatile integrals. Integral transforms, known as $P$-transforms, are also associated with the integrals in (4.8), see for example [18],[19].

  \vskip.3cm\noindent{\bf 4.3.\hskip.3cm Fractional calculus models}

  \vskip.3cm In a series of papers the authors ([12], [13], [14], [15], [17], [22], [28]) have shown recently that fractional integrals can be classified into the forms in (4.2) and (4.6) or fractional integral operators of the second kind or right-sided fractional integral operators can be considered as Mellin convolution of a product as in (4.2) and left-sided or fractional integral operators of the first kind can be considered as Mellin convolution of a ratio where the functions $f_1$ and $f_2$ are of the following forms:
  $$f_1(x)=\phi_1(x)(1-x)^{\alpha-1},0\le x\le 1, f_2(y)=\phi_2(y)f(y)\eqno(4.9)
  $$where $\phi_1$ and $\phi_2$ are pre-fixed functions, $f(y)$ is arbitrary and $f_1(x)=0$ outside the interval $0\le x\le 1$. Thus, essentially, all fractional integral operators belong to the categories of Mellin convolution of a product or ratio where one function is a multiple of type-1 beta form and the other is arbitrary. The right-sided or type-2 fractional integral of order $\alpha$ is denoted by $D_{2,u}^{-\alpha}f$ and defined as

  $$D_{2,u}^{-\alpha}f=\int_v\frac{1}{v}f_1(\frac{u}{v})f_2(v){\rm d}v\eqno(4.10)
  $$and the left-sided or type-1 fractional integral of order $\alpha$ is given by
  $$D_{1,u}^{-\alpha}f=\int_v\frac{v}{u^2}f_1(\frac{v}{u})f_2(v){\rm d}v\eqno(4.11)
  $$where $f_1$ and $f_2$ are as given in (4.9). Let $n$ be a positive integer such that $\Re(n-\alpha)>0$. The smallest such $n$ is $[\Re(\alpha)]+1=m$ where $[\Re(\alpha)]$ denotes the integer part of $\Re(\alpha)$. Here $D_{2,u}^{-\alpha}f$ and $D_{1,u}^{-\alpha}f$ are defined as in (4.10) and (4.11) respectively. Let $D=\frac{{\rm d}}{{\rm d}u}$ the ordinary derivative with respect to $u$ and $D^n$ be the $n$-th order derivative. Then the fractional derivative of order $\alpha$ is defined as

  $$D^{\alpha}f=D^n[D_{i,u}^{-(n-\alpha)}f] \mbox{  in the Riemann-Liouville sense and}
  $$
  $$D^{\alpha}f=[D_{i,u}^{-(n-\alpha)}D^n f] \mbox{  in the Caputo sense}\eqno(4.12)
  $$for $i=1,2$, see also [31].
  \vskip.2cm The input-output model that we started with, when applied to reaction-diffusion problems can result in fractional order reaction-diffusion differential equations. Such fractional order differential equations are seen to provide solutions which are more relevant to practical situations compared to the solutions coming from differential equations in the conventional sense or involving integer-order derivatives. Some of the relevant papers in this direction may be seen from [12], [13], [16], [17].

  \vskip.3cm\noindent{\bf Acknowledgement}
  \vskip.3cm The authors would like to thank the Department of Science and Technology, Government of India, for the financial assistance for this work under Project Number SR/S4/MS:287/05 and the Centre for Mathematical Sciences India for facilities.

  \vskip.3cm\noindent{\bf References}

  \vskip.3cm\noindent[1]~~W.J. Anderson, H.J. Haubold and A.M. Mathai (1994): Astrophysical thermonuclear functions, {\it Astrophysics and Space Science}, {\bf 214(1-2)}, 49-70.
  \vskip.2cm\noindent[2]~~H.J. Haubold and A.M. Mathai (1984): On the nuclear energy generation rate in a simple analytic stellar model, {\it Annalen der Physik}, {\bf 41}, 372-379.
  \vskip.2cm\noindent[3]~~H.J. Haubold and A.M. Mathai (1984): On nuclear reaction rate theory, {\it Annalen der Physik}, {\bf 41}, 380-396.
  \vskip.2cm\noindent[4]~~H.J. Haubold and A.M. Mathai (1988): {\it Modern Problems in Nuclear and Neutrino Astrophysics}, Akademie-Verlag, Berlin.
  \vskip.2cm\noindent[5]~~K. Sakurai (2014): {\it Solar Neutrino Problems - How They Were Solved}, TERRAPUB, Tokyo.
  \vskip.2cm\noindent[6]~~H.J. Haubold and A.M. Mathai (1995): A heuristic remark on the periodic variation in the number of solar neutrinos detected on Earth, {\it Astrophysics and Space science}, {\bf 228}, 113-134.
  \vskip.2cm\noindent[7]~~H.J. Haubold, A.M. Mathai, and R.K. Saxena (2014) Analysis of solar neutrino data from Super-Kamiokande I and II, {\it Entropy}, {\bf 16}, 1414.
\vskip.2cm\noindent[8]~~A.M. Mathai and H.J. Haubold (2013) On a generalized entropy measure leading to the pathway model with a preliminary application to solar neutrino data, {\it Entropy}, {\bf 15}, 4011.
   \vskip.2cm\noindent[9]~~H.J. Haubold and A.M. Mathai (1998): On thermonuclear reaction rates, {\it Astrophysics and Space science}, {\bf 258}, 185-189.
   \vskip.2cm\noindent[11]~~H.J. Haubold and A.M. Mathai (1998): An integral arising frequently in astronomy and physics, {\it SIAM Review}, {\bf 40(4)}, 995-997.
   \vskip.2cm\noindent[12]~~H.J. Haubold and A.M. Mathai (2000): The fractional kinetic equation and thermonuclear functions, {\it Astrophysics and Space Science}, {\bf 273(1-4)}, 53-63.
   \vskip.2cm\noindent[13]~~H.J. Haubold and A.M. Mathai (2002): On fractional kinetic equations, {\it Astrophysics and Space Science}, {\bf 282}, 281-287.
   \vskip.2cm\noindent[14]~~H.J. Haubold and A.M. Mathai (2006): A certain class of Laplace transform with application to reaction and reaction-diffusion equations, {\it Astrophysics and Space Science}, {\bf 305}, 283-288.
   \vskip.2cm\noindent[15]~~H.J. Haubold and A.M. Mathai (2007): Pathway model, superstatistics, Tsallis statistics and a measure of entropy, {\it Physica A}, {\bf 375}, 110-122.
   \vskip.2cm\noindent[16]~~H.J. Haubold, A.M. Mathai and R.K. Saxena (2010): Solution of certain fractional kinetic equations and a fractional diffusion equation, {\it Journal of Mathematical Physics}, {\bf 51}, 103506-1,103506-8.
   \vskip.2cm\noindent[17]~~H.J. Haubold, A.M. Mathai and R.K. Saxena (2011): Further solutions of fractional reaction-diffusion equations in tems of the H-function, {\it Journal of Computational and Applied Mathematics}, {\bf 235}, 1311-1316.
   \vskip.2cm\noindent[18]~~D. Kumar and A.A. Kilbas (2010): Fractional calculus of P-transform, {]\it Fractional Calculus and Applied Analysis}, {\bf 13(3)}, 317-328.
   \vskip.2cm\noindent[19]~~D. Kumar (2011): P-transform, {\it Integral Transforms and Special Functions}, {\bf 22(8)}, 603-3611.
   \vskip.2cm\noindent[20]~~A.M. Mathai (1989): On a system of differential equations connected with the gravitational instability in a multi-component medium in Newtonian cosmology, {\it Studies in Applied Mathematics}, {\bf 80}, 75-03.
   \vskip.2cm\noindent[21]~~A.M. Mathai (1993): The residual effect of a growth-decay mechanism and the covariance structures, {\it The Canadian Journal of Statistics}, {\bf 21(3)}, 277-283.
   \vskip.2cm\noindent[22]~~A.M. Mathai (2009): Fractional integrals in the matrix-variate cases and connection to statistical distributions, {\it Integral Transforms and Special Functions}, {\bf 20(12)}, 871-882.
   \vskip.2cm\noindent[23]~~A.M. Mathai (2005): A pathway to matrix-variate gamma and normal densities, {\it Linear Algebra and Its Applications}, {\bf 396}, 317-328.
   \vskip.2cm\noindent[24]~~A.M. Mathai (2012): Generalized Kr\"atzel integrals associated with statistical densities, {\it International Journal of Mathematical Analysis}, {\bf 6(51)}, 2501-2510.
   \vskip.2cm\noindent[25]~~A.M. Mathai (2012): Stochastic models under power transformations and exponentiation, {\it Journal of the Indian Society for Probability and Statistics}, {\bf 13}, 1-19.
   \vskip.2cm\noindent[26]~~A.M. Mathai and H.J. Haubold (2007): On generalized entropy measure and pathways, {\it Physica A}, {\bf 385}, 493-500.
   \vskip.2cm\noindent[27]~~A.M. Mathai and H.J. Haubold (2008): Pathway parameter and thermonuclear functions, {\it Physica A}, {\bf 387}, 2462-2470.
   \vskip.2cm\noindent[28]~~A.M. Mathai and H.J. Haubold (2008): {\it Special Functions for Applied Scientists}, Springer, New York.
   \vskip.2cm\noindent[29]~~A.M. Mathai and H.J. Haubold (2011): Mittag-Leffler functions to pathway model to Tsallis statistics, {\it Integral Transforms and Special Functions}, {\bf 21(11)}, 867-875.
   \vskip.2cm\noindent[30]~~A.M. Mathai and H.J. Haubold (2011): A pathway for Bayesian statistical analysis to superstatistics, {\it Applied Mathematics and Computations}, {\bf 218}, 799-804.
   \vskip.2cm\noindent[31]~~A.M. Mathai and H.J. Haubold (2013): Erdelyi-Kober fractional integral operators from a statistical perspective I-IV, {\it arXiv:1303.3978-3981}.
   \vskip.2cm\noindent[32]~~A.M. Mathai, R.K. Saxena and  H.J. Haubold (2010): {\it The H-function: Theory and Applications}, Springer, New York.
   \vskip.2cm\noindent[33]~~A.M. Mathai and H.J. Haubold (2011): A versatile integral in physics and astronomy, {\it arXiv:1109.5173}.

  \end{document}